# Giant peak effect observed in an ultrapure YBa$_2$Cu$_3$O$_{6.993}$ crystal

Jing Shi* and X. S. Ling†
*Department of Physics, Brown University, Providence, Rhode Island 02912*

Ruixing Liang, D. A. Bonn, and W. N. Hardy
*Department of Physics and Astronomy, University of British Columbia, Vancouver, British Columbia, Canada V6T 1Z1*

A giant peak in the temperature dependence of the screening current is observed in the ac magnetic response of an ultrapure YBa$_2$Cu$_3$O$_{6.993}$ crystal in a magnetic field. At $H=2.0$ T ($H\|\mathbf{c}$), the screening current density $J_c(T)$ exhibits a 35-fold rise with a 0.5 K increase in temperature, indicating an abrupt $10^3$-fold collapse in the characteristic volume of ordered regions in the vortex array. The peak-effect anomaly is most pronounced for $H<4.0$ T, but detectable up to 7.0 T. The temperature dependence of the equilibrium magnetization exhibits a small discontinuous jump (for high fields) inside the peak-effect regime, suggesting that the underlying phase transition is a weak first-order vortex-lattice melting transition. [S0163-1829(99)51042-4]

One of the unsolved puzzles in high-$T_c$ superconductors is that the critical current density $J_c(T)$ of YBa$_2$Cu$_3$O$_{7-\delta}$ (YBCO) crystals in a magnetic field, instead of decreasing monotonically with increasing temperature, sometimes makes a surprising upturn before it vanishes.[1–6] This anomalous peak effect is believed to be due to a vanishing shear modulus of the vortex lattice which in turn leads to more effective pinning of vortex lines by random impurities. Since a finite shear modulus is the defining property of a solid, and its disappearance is the hallmark of solid melting, it has been suggested that the peak effect is either caused by or related to an underlying vortex-lattice melting transition.[7–9]

For regular (crystalline) solid melting, the vanishing of the shear modulus at the melting point is accompanied by thermodynamic signals such as a discontinuous change in density. There is compelling thermodynamic evidence, e.g., a jump in the equilibrium magnetization $M(T)$,[10–12] suggesting that such a discontinuous change in density does occur in the vortex state in high-quality YBCO crystals. For the peak effect and the magnetization jump to be the corresponding signatures, i.e., vanishing shear modulus and discontinuous change in density of vortex solid melting, they should occur at the same temperature, or in close vicinity. Where exactly the peak effect occurs relative to the $M(T)$ transition on the field-temperature ($H$-$T$) phase diagram is crucial for understanding the peak effect and the underlying phase transition in high-$T_c$ superconductors.

Unfortunately, for reasons that are still not well understood, rarely can one observe a peak effect and an $M(T)$ transition simultaneously. In a recent work[6] in which both effects were detected in the same untwinned YBCO crystal, at $H=1.0$ T ($H\|\mathbf{c}$), the peak of the peak effect does coincide with the jump in $M(T)$. However at higher fields ($>1.5$ T) where the $M(T)$ jumps are pronounced, $J_c(T)$ decreases monotonically with $T$ across the transition[6] (also see Ref. 13).

The main difficulty, however, in accepting a phase transition scenario for the peak effect, is that the magnitudes of the observed $J_c(T)$ peaks are too small. For the observed $J_c(T)$ peaks in YBCO crystals,[1–6] the ratios of peak-$J_c(T_p)$-to-onset-$J_c(T_o)$ were typically less than 2. In the context of collective pinning theory,[14] in which $J_c(T)$ is determined by the collective volume $V_c$, $J_cB=\sqrt{nf_p^2/V_c}$, where $n$ and $f_p$ are respectively the density and characteristic force of the pinning centers, a factor of 2 enhancement in $J_c(T)$ indicates a mere factor of 4 decrease in $V_c$, a rather small loss of order for a disordering phase transition.

In this paper, we report a striking observation of a giant peak effect in $J_c(T)$ in the vortex state of an ultrapure YBa$_2$Cu$_3$O$_{6.993}$ crystal, made possible by the recent work in the crystal growth of high-$T_c$ oxide superconductors using BaZrO$_3$ crucibles. In an applied field of 2.0 T, $H\|\mathbf{c}$, $J_c(T)$ shows a 35-fold increase from the onset ($T_o$) to the peak ($T_p$) with $T_p-T_o<0.5$ K. Thermodynamic signatures in the equilibrium magnetization $M(T)$ are also found in the peak-effect regime, suggesting that the underlying phase transition is the vortex-lattice melting.

The sample is an ultrapure YBa$_2$Cu$_3$O$_{6.993}$ (overdoped) crystal grown in a bulk BaZrO$_3$ crucible. The growth procedures for ultrahigh purity YBCO crystals using highly inert BaZrO$_3$ crucibles are now well documented.[15,16] The chemical and structural characterization of these crystals confirmed that they have very low levels of impurity elements and a high degree of crystalline order.[16] As an example, the full width at half maximum (FWHM) of the x-ray rocking curve (006) from this sample is less than 0.006°, a notably small value for YBCO crystals. The sample is a perfect rectangle with dimensions $1.53\times1.28\times0.065$ mm$^3$ and the **c** axis along the shortest dimension. The bulk of the crystal is twin free except for a single twin boundary cutting the very tip of one of the four corners, forming a triangle which has an area less than 0.12% of the total sample area. No measurable effect is expected from this single twin on the ac response which probes the macroscopic screening current in the sample. The ac susceptibility at zero dc field, $H_{ac}=25.0$ Oe, and frequency $f=100.0$ kHz, shows a single step [width ($10-90\%)=0.85$ K] in the real part $4\pi\chi'(T)$ and a single peak (at $T_c=88.60$ K) in the imaginary part $4\pi\chi''(T)$. The dc magnetization measurement, using a commercial super-conducting quantum interference device (SQUID) magnetometer (Quantum Design), at 3.0 Oe (field cooled) gives a transition temperature $T_c=88.55$ K and width (10–90 %) $\Delta T_c=0.55$ K.

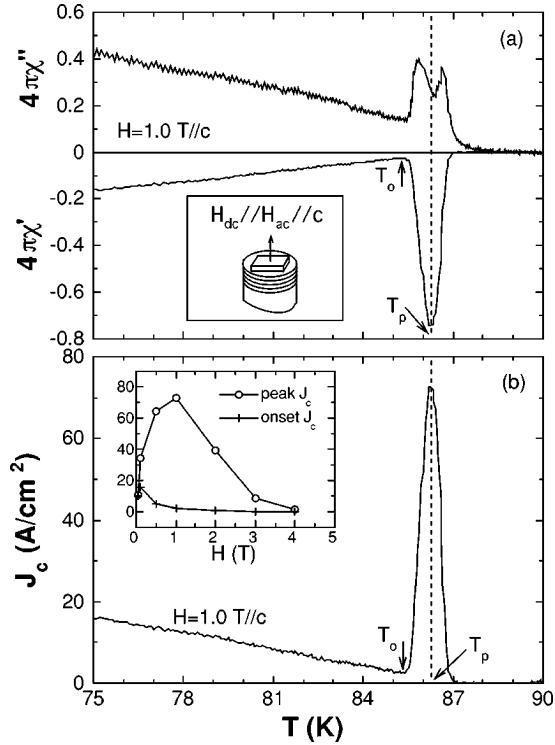

FIG. 1. (a) $4\pi\chi'(T)$ and $4\pi\chi''(T)$ at $H_{dc}=1.0$ T. $H_{ac}=25.0$ Oe and $f=100.0$ kHz. Inset: the experimental configuration. (b) $J_c(T)$. Inset: the peak $J_c(T_p)$ and onset $J_c(T_o)$ at different magnetic fields. $J_c(T)$ is calculated using a simplified critical-state model (applicable when the ac fields are large enough to penetrate to the center of the sample), $4\pi\chi'=-\pi J_c d/H_{ac}$ (in cgs), where $d$ is chosen as the sample thickness instead of the width (see Ref. 1 and Refs. 19 and 20 below).

The real part of nonlinear ac susceptibility is a measure of the screening current induced in the superconductor by the ac magnetic field. The peak effect is identified as a dip in its temperature dependence.[1,2,4,6] The basic setup, as shown in the inset of Fig. 1(a), consists of a 50-turn coil, wound using a fine insulated copper wire, on the end of a sapphire rod (2.0 mm in diameter) on which the crystal is glued with Apiezon-N grease. The impedance of the coil is measured using a two-phase SR-844 lockin amplifier, and the coil inductance and resistance are related to the ac susceptibility of the sample by $L=L_0(1+4\pi\chi'\zeta)$ and $R=R_0+4\pi\chi''\omega L_0\zeta$, where $R_0$ and $L_0$ are the resistance and the inductance of the empty coil, $\zeta$ is the coil-sample coupling constant (filling factor), and $\omega=2\pi f$. $H_{ac}$ is held constant for each temperature sweep by using a constant amplitude ac current source (HP3245A). The details of this method have been described elsewhere.[1]

Figure 1(a) shows the temperature dependence of the real $4\pi\chi'(T)$ and the imaginary $4\pi\chi''(T)$ parts of the ac susceptibility at $H=1.0$ T. The sample-coil coupling factor $\zeta$ ($=0.026$) is determined by choosing $4\pi\chi'(70\text{ K},0\text{ T})=-1.0$. The large dip in $4\pi\chi'(T)$ of the ac susceptibility corresponds to a giant peak in the temperature dependence of the screening current in the sample. At low frequencies and at an ac field large enough to penetrate deep into the bulk, the screening current in the vortex state is the critical current supported by pinning centers in the bulk.[17] In this ultrapure sample, we expect the oxygen vacancies to be the primary pinning centers. The critical current density $J_c(T)$, shown in Fig. 1(b), is calculated from the susceptibility data using a critical-state model.[17] Since the dependence of $J_c$ on the ac field is neglected in the model used here, small corrections to Fig. 1(b) are expected if one uses more refined treatments.[18–20] The order of magnitude and the temperature dependence of $J_c$ are expected to hold.

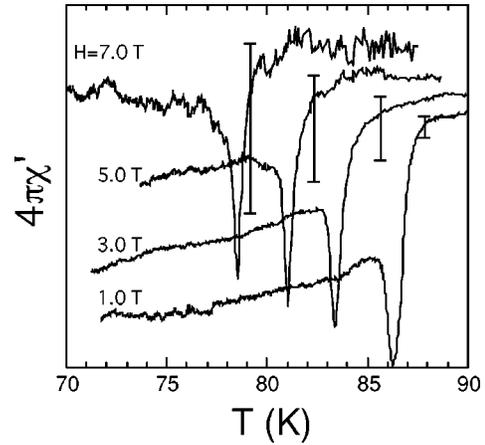

FIG. 2. $4\pi\chi'(T)$ for $f=1.0$ MHz. $H_{dc}=7.0$, 5.0, 3.0, and 1.0 T (from left to right), and $H_{ac}=25.0$ Oe. $H\|\mathbf{c}$. The four scale bars for each field are equal to $4\pi\chi'=-0.07$.

The peak effect is most pronounced in the intermediate field range 0.1–2.0 T, over which a peak effect in YBCO has been frequently reported.[1–6] At 1.0 T, from the onset to the peak, over a temperature interval of 0.8 K, $J_c(T)$ rises by a factor of 31. The ratio of the peak $J_c$ to the onset $J_c$ is 35 for 2.0 T, which indicates a more than $10^3$-fold drop in the collective volume $V_c$ of the pinned vortex array.

As shown in the inset of Fig. 1(b), the $J_c$ peak value has a maximum at 1.0 T and the $J_c$ onset value decreases with increasing field. The peak $J_c$ and onset $J_c$ merge at very low fields. At 4.0 T, the onset $J_c$ drops below our detection level ($\approx 0.5$ A/cm$^2$) while the peak is still detectable (height $\approx 1.5$ A/cm$^2$, width$\approx 0.4$ K, data not shown). At above 4.0 T, the $\chi'(T)$ dip is no longer discernible. However, the way in which the $\chi'(T)$ dip disappears at high fields in this sample is very different from what was observed previously.[4,6] In the earlier samples, the $\chi'$ dip broadens and then turns into a single step in $\chi'(T)$. Here, not just the $\chi'$ dip, but no signal from the screening current at all can be detected in the temperature range (70–90 K). We attribute the absence of a detectable signal at $f=100.0$ kHz (for $H>4.0$ T) to the decay of the screening current caused by thermally activated depinning, in addition to the ultrahigh purity (thus the very weak disorder) of the crystal. Since the decay of the screening current should be less severe when probed at a short time scale, one might expect to see the peak effect again in the ac susceptibility if measured at a higher frequency. This is indeed true. Figure 2 shows $\chi'(T)$ at $H$ $=1.0$, 3.0, 5.0, and 7.0 T, for $f=1.0$ MHz and $H_{ac}=25.0$ Oe. Up to the highest field (7.0 T) of our system, a $\chi'(T)$ dip remains observable.

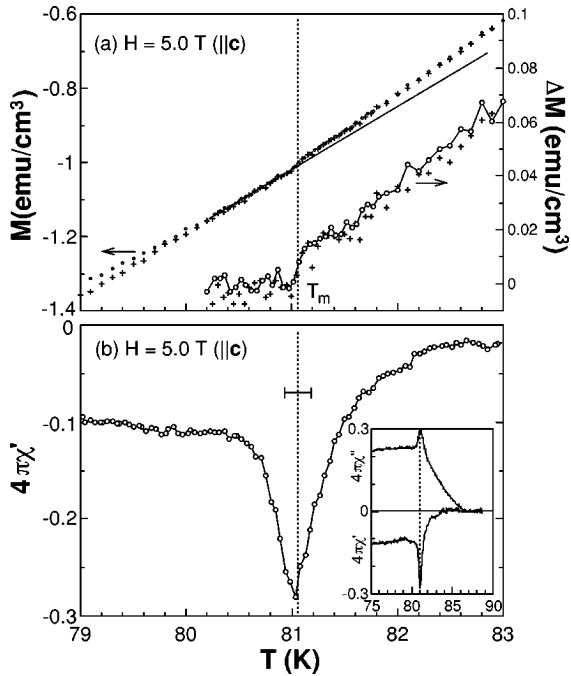

FIG. 3. (a) $M(T)$ (ZFC: crosses; FC: circles) at $H=5.0$ T ($\mathbf{H}\|\mathbf{c}$). $\Delta M(T)$ is the difference between the reversible part of the data ($T>80.3$ K) and the solid line. (b) $4\pi\chi'(T)$ at $f=1.0$ MHz and $H_{ac}=25.0$ Oe. Inset: $4\pi\chi'(T)$ and $4\pi\chi''(T)$ over a wider temperature range. The dashed lines indicate $T_m$.

The detection of a $\chi'(T)$ dip in the high-frequency ac response made possible the comparison of the peak effect and the dc magnetization jump, since the latter can only be detected at high fields in this sample. Figure 3(a) shows the temperature dependence of the dc magnetization $M(T)$ of the sample at $H=5.0$ T. The crosses are for the zero-field-cooled (ZFC) measurements (the sample is cooled in zero field and the magnetic field is turned on at $\approx 70.0$ K) and the circles are data taken as the sample is cooled in the field (FC). For $T>80.3$ K, $M(T)$ is essentially reversible, thus is the equilibrium magnetization $M(T)$. A small, but sharp, jump in $M(T)$ can be identified at $T_m=81.06$ K, as shown by $\Delta M(T)$ in Fig. 3(a). The width of the jump is $\Delta T_m \approx 0.08$ K.

For comparison, the $4\pi\chi'(T)$ data at $H=5.0$ T and $f=1.0$ MHz are replotted in Fig. 3(b). The $\chi'(T)$ dip is at $T_p=81.03$ K. We should point out that the dc magnetization and the ac susceptibility are measured in two separate setups, so that there is a relative uncertainty of 0.25 K between the temperature readings in the two systems in this temperature range, as indicated by the error bar in Fig. 3(b). However, given this uncertainty, it is clear that the $M(T)$ jump at $H=5.0$ T is well within the temperature regime of the peak effect.

Figure 4(a) shows $M(T)$ at $H=3.0$ T. No jump can be identified in the data; instead, a clear change of slope is observed in $M(T)$ (see inset). The dashed line indicates the transition temperature defined by the crossing point of the two linear parts of $M(T)$ (by extrapolation). In the transition region, there is a bump in $M(T)$. (We note that the SQUID validation curves show a small deformation in this region, thus the bump could be an artifact caused by the reappearance of strong vortex pinning due to the peak effect.) In the same transition region, a sharp dip in $4\pi\chi'(T)$ appears, even at low frequencies. Figure 4(b) shows $4\pi\chi'(T)$ at $f=100.0$ kHz, and a replot for $f=1.0$ MHz. Note that the data for $f=100.0$ kHz have been expanded by a factor of 4. Again, given the uncertainty in the relative temperature readings [the bar in Fig. 4(b)], the peak effect and the $M(T)$ transition, the change of slope, occur at the same temperature.

Figure 5 shows the $H$-$T$ phase diagram constructed using the $\chi'(T)$ dip temperature $T_p$ at 1.0 MHz for different mag-netic fields. As shown in Fig. 4(b), $T_p$ changes very little for the frequency range 0.1–1.0 MHz. The jump and the change of slope in the equilibrium magnetization $M(T)$ coincide, within our experimental error, with $T_p$. In an earlier report,[6] it was also found that, although only demonstrated for $H = 1.0$ T, the peak of the peak effect occurs at exactly the same temperature as the dc magnetization jump. The solid line in Fig. 5 is a fit to a melting line predicted[21] for three-dimensional (3D) $XY$ fluctuations $H = H_0(1 - T/T_c)^n$ (T), where the fitting parameter $T_c = 88.60$ K is consistent with the measured value, and $H_0 = 133.0$ T and $n = 1.33$. These parameters are consistent with those determined from a calorimetric measurement[22] of a similar ultrapure overdoped $YBa_2Cu_3O_{7.00}$ crystal, for which $T_c = 87.8$ K, $H_0 = 135$ T, and $n = 1.33$. An exponent of $n = 2.0$ also gives a reasonable fit but the fitted $T_c$ is much larger than the measured value.

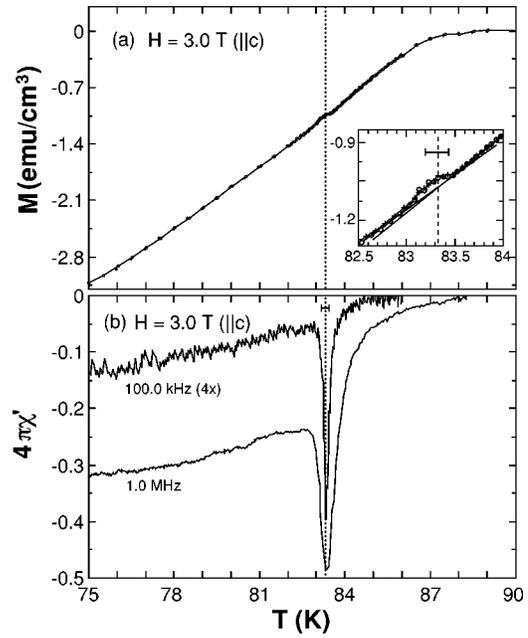

FIG. 4. (a) $M(T)$ (ZFC: crosses; FC: circles) at $H=3.0$ T ($\mathbf{H}\|\mathbf{c}$). Inset: The expanded view of $M(T)$ in the transition region. The straight lines are drawn by hand through the linear parts of $M(T)$. (b) $4\pi\chi'(T)$ at $f=100.0$ kHz and 1.0 MHz, $H_{ac}=25.0$ Oe.

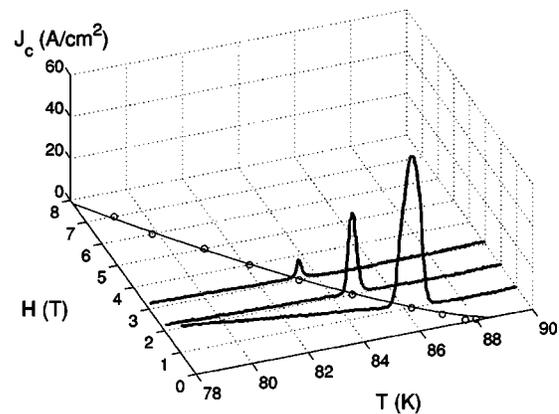

FIG. 5. A 3D phase diagram for the peak effect in $YBa_2Cu_3O_{6.993}$. The open circles show the location ($T_p$) of the peak effect for both 100.0 kHz and 1.0 MHz. The solid line on the $H$-$T$ plane is a fit to the 3D $XY$ melting conjecture (see text). The $J_c$ peaks are shown for $H=1.0$, 2.0, and 3.0 T measured at 100.0 kHz.

Our main conclusion is that the peak effect in the screening current and the jump or the change of slope in $M(T)$ can indeed be interpreted as the signatures, i.e., vanishing shear modulus and a change of density, of an underlying vortex-lattice melting transition. However, presently there is no complete explanation for all of our data. At $H = 5.0$ T, the width of the $\chi'(T)$ dip at 1.0 MHz is about 1.0 K while the width of the jump in $M(T)$ is only 0.08 K. We point out that the large width of the $\chi'(T)$ dip cannot be due to a smearing of the phase transition by the ac field. For $H_{ac} = 25.0$ Oe, the smearing only amounts to $\approx 0.004$ K at 5.0 T, and 0.006 K at 1.0 T, due to the large slope of the phase boundary on the $H$-$T$ phase diagram. One possible interpretation is that the vortex-lattice melting/freezing is *weakly* first order. The large width of the peak effect [$\chi'(T)$ dip] suggests that significant disordering (as the transition is approached from below), or ordering (when the transition is approached from above) processes have taken place in the vortex array prior to the melting/freezing point.

One of the authors (X.S.L.) wishes to thank Professor Leon Cooper for his interest in this work; S. Bhattacharya, A. Houghton, J. M. Kosterlitz, M. C. Marchetti, R. A. Pelcovits, and E. Zeldov for discussions; J. M. Valles for a critical reading of the manuscript; and the A.P. Sloan Foundation and the NSF-DMR for support. The work at UBC was supported by the NSERC and the CIAR of Canada.